\documentclass[aps,prc,preprint]{revtex4-1}

\usepackage{graphicx} 
\usepackage{colordvi}
\begin{document}

\title{Examination of scaling of Hanbury-Brown--Twiss radii with charged
particle multiplicity}

\author{Gunnar Gr\"af}
\email{graef@th.physik.uni-frankfurt.de}
\affiliation{Institut f\"ur Theoretische Physik, Goethe Universit\"at Frankfurt,Germany}
\affiliation{Frankfurt Institute for Advanced Studies (FIAS), Ruth-Moufang-Str. 1, 60438 Frankfurt, Germany}
\author{Qingfeng Li}
\affiliation{School of Science, Huzhou Teachers College, Huzhou 313000, P.R. China}
\author{Marcus Bleicher}
\affiliation{Institut f\"ur Theoretische Physik, Goethe Universit\"at Frankfurt,Germany}
\affiliation{Frankfurt Institute for Advanced Studies (FIAS), Ruth-Moufang-Str. 1, 60438 Frankfurt, Germany}

\begin{abstract} In the light of the recent LHC data on proton-proton and
lead-lead collisions we examine the question of the multiplicity scaling of HBT
radii in relativistic nuclei and particle interactions. Within the UrQMD
transport approach we study a large variety of system sizes at different beam
energies and extract the HBT radii. In the calculation, we find a good scaling
of the radii as a function of charged particle multiplicity, if the change in
the multiplicity is caused by a change of centrality at the same energy.
However, the scaling is only approximate when the energy, $\sqrt{s}$, is
changed and breaks down when comparing pp to AA reactions.
\end{abstract} \maketitle

\section{Introduction}

The properties of strongly interacting matter are described by the theory of
Quantum-Chromo-Dynamics (QCD). To explore the details of QCD matter under
extreme conditions, one needs to compress and heat up QCD matter to regimes
present microseconds after the Big Bang. Today these conditions can only be
found in the interior of neutron stars or created in heavy-ion collisions at
relativistic energies. Over the last decade the experimental programs at the
SPS (e.g. with the NA49, CERES and NA50/NA60 experiments) and at RHIC (e.g.
PHENIX, STAR, PHOBOS, and BRAHMS) have provided exciting pioneering data on the
equation of state, the transport properties of the matter created and its
spatial distributions
\cite{Adams:2004yc,Lisa:2000no,Alt:2007uj,Afanasiev:2002mx,Adamova:2002wi,Abelev:2009tp,Back:2004ug,Back:2005hs,Back:2002wb,Abelev:2008ez}.
These programs are currently extended into a system size scan with NA61 at SPS
and a systematic beam energy scan with the RHIC-BES initiative. In addition, at
the high energy frontier unprecedented data from the Large Hadron Collider
(LHC) for (high multiplicity) proton-proton and Pb+Pb reactions up to
$\sqrt{s_{NN}}=7$~TeV has become available (see
\cite{Aamodt:2011kd,Aamodt:2011mr} for HBT related results). Particle
correlations, i.e. Hanbury-Brown Twiss correlation (HBT) or femtoscopy allow
to gain deeper insights into the emission patterns and coherence regions of the
matter created \cite{Pratt:1984su,Sinyukov:1989xz,Hama:1987xv,nucl-th/9901094}.
One generally assumes that the observed HBT radii scale with the charged
particle density (or number of participants) as the charged particle density
should be a good proxy for the final state volume \cite{Lisa:2005dd}. However,
the interferometry volume may not only depend on multiplicity, but also on the
initial size of the colliding system \cite{Sinyukov:2011mw}. Indeed, one of the
surprising LHC results concerns the scaling violation observed in pp reactions
as compared to AA reactions at lower energies at the same charged particle
density. In this paper, we want to explore the spatial structure of the source
created in collisions of various heavy ions at different energies and
centralities to shed light on the observed scaling violation when going from
proton-proton to AA collisions at the LHC. Other investigations on the charged
particle yield scaling can be found in 
\cite{Adamova:2002wi,Adamova:2002ff,Akkelin:2003kp,Akkelin:2005ms}. Results for
PbPb and pp reactions at the LHC within the same model can be cound in
\cite{Graef:2012za,Li:2012ta}.

\section{Model and HBT calculation}

For the present study we employ the UrQMD \cite{Bass:1998ca,Bleicher:1999xi}
transport model in version 3.3 (for details of version 3.3 see
\cite{Petersen:2008kb,Petersen:2008dd}). The model can be downloaded from
\cite{urqmd-webpage}. For earlier HBT results from UrQMD see
\cite{Li:2006gp,Li:2007yd,Li:2008qm,Li:2008ge}. UrQMD is
a microscopic non-equilibrium transport model. It models the space-time
evolution of nucleus-nucleus collisions from the beginning of the collision
until the kinetic freeze-out. Particles are produced via hard collisions,
string excitation and fragmentation and via resonance excitation and decay.\\

For the calculation of the HBT radii we use the pion freeze-out distribution
from UrQMD. Then we calculate the HBT correlation function by
\cite{nucl-th/9901094,Lisa:2005dd}
\begin{equation}
  C({\bf q},{\bf K}) = 1 + \int d^4x\ cos(q\cdot x)\  d(x,K)\ , \label{eqn:correlation}
\end{equation}
where $C$ is the correlation function, $q$ is the four-momentum distance of the correlated particles, $K=(p_1+p_2)/2$ is the pair momentum, $x$ is the particle separation four-vector and $d$ is the normalized pion freeze-out separation distribution, which is an even function of $x$. For the analysis in this paper all values are taken in the pair longitudinal comoving system (LCMS). Since UrQMD generates a discrete set of freeze-out points, the integral in Eq. \ref{eqn:correlation} is substituted by a sum.\\

The HBT radii $R_{ij}$ are obtained by fitting the function
\begin{equation}
  C({\bf q},{\bf K}) = 1 + \lambda({\bf K}) exp \left [ - \sum_{i,j=o,s,l} q_iq_jR_{ij}^2({\bf K}) \right ]
\end{equation}
to the calculated three-dimensional correlation functions. For the analysis in
this paper the correlation functions are fitted over a range $|{\rm q_i}| < $
800 MeV/c  for proton-proton collisions, $|{\rm q_i}| < $ 300 MeV/c for
carbon-carbon collisions and $|{\rm q_i}| < $ 150 MeV/c for all other
collisions. The difference in the momentum ranges is motivated by the fact
that the width of the peak in the correlation function gets broader for smaller
systems. Thus, the fit range is bigger for proton-proton and carbon-carbon, than
it is for lead-lead collisions.

\section{Scaling of the HBT radii}

\begin{figure}
\center
\includegraphics[width=.9\textwidth]{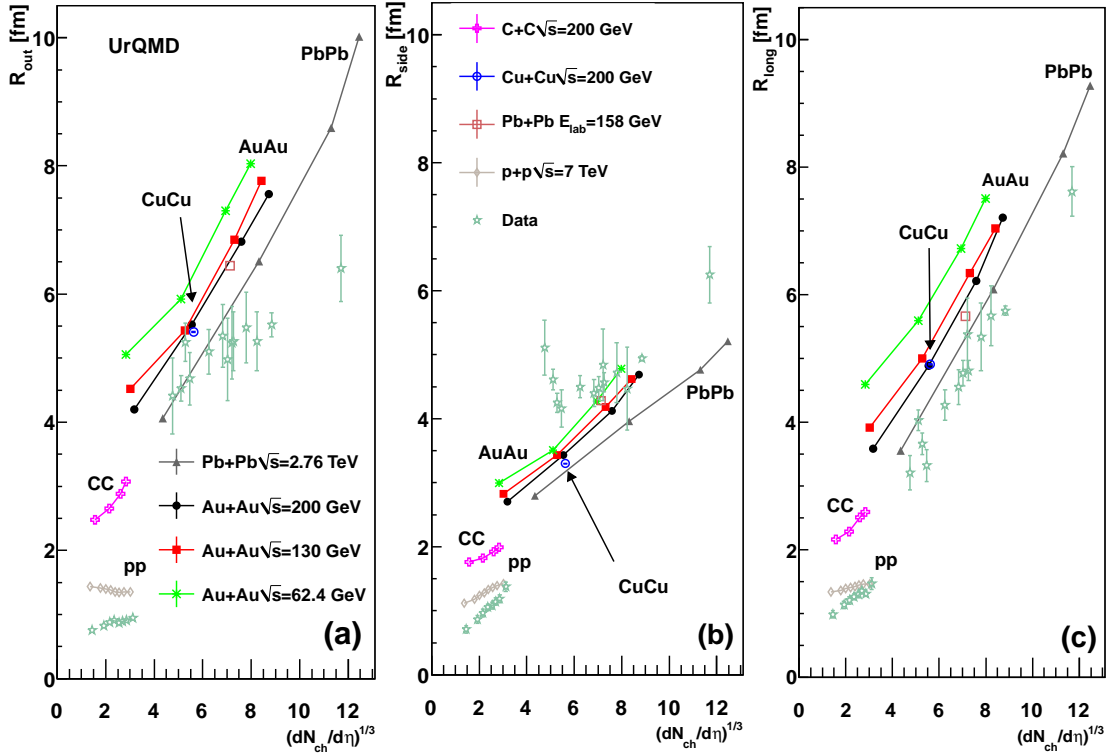}
\caption{(Color online) The three HBT radii $R_{out}$, $R_{side}$ and $R_{long}$ as a function
of the charged particle multiplicity at midrapidity, $(dN_{ch}/d\eta)^{1/3}$
and fixed $k_T=300-400$~MeV. The lines with symbols are the simulation results.
The gray triangles, the black circles, the red squares and the green crosses
are for lead-lead collisions at $\sqrt{s}=$ 2760, 200, 130, 62.4 GeV (in the
same order) at 0-5\%, 5-20\%, 20-50\% and 50-80\% centrality for the different
points. The pink crosses are results for carbon-carbon at $\sqrt{s}=$ 200 GeV
for the same centrality classes and the beige diamonds show results for various
multiplicity classes from proton-proton collisions \cite{Graef:2012za}. Blue
circles and brown squares depict results for central copper-copper events at
$\sqrt{s}=$ 200 GeV and central lead-lead collisions at $E_{lab}=$ 158 GeV. The
green stars are experimental results for central gold and lead collisions at
$k_T =$ 300 GeV/c taken from
\cite{Adams:2004yc,Lisa:2000no,Alt:2007uj,Afanasiev:2002mx,Adamova:2002wi,Abelev:2009tp,Back:2004ug,Back:2005hs,Back:2002wb,Abelev:2008ez,Aamodt:2011kd,Aamodt:2011mr}.
\label{fig:radii}}
\end{figure}

Fig. \ref{fig:radii} shows the three HBT radii $R_{out}$, $R_{side}$ and
$R_{long}$ as a function of the charged particle multiplicity at midrapidity
($|\eta|<1.2$ for pp and $|\eta|<0.8$ for all other classes),
$(dN_{ch}/d\eta)^{1/3}$ and fixed $k_T=300-400$~MeV. The lines with symbols are
simulation results for lead-lead collisions at $\sqrt{s}=$ 2760, 200, 130, 62.4
GeV for 0-5\%, 5-20\%, 20-50\% and 50-80\% centrality, for carbon-carbon at
$\sqrt{s}=$ 200 GeV in the same centrality classes, for proton-proton at
$\sqrt{s}=$ 7 TeV with different $dN_{ch}/d\eta$ classes, for central
copper-copper collisions at $\sqrt{s}=$ 200 GeV and for central lead-lead
collisions at $E_{lab}=$ 158 GeV. The green stars are experimental results
taken from
\cite{Adams:2004yc,Lisa:2000no,Alt:2007uj,Afanasiev:2002mx,Adamova:2002wi,Abelev:2009tp,Back:2004ug,Back:2005hs,Back:2002wb,Abelev:2008ez,Aamodt:2011kd,Aamodt:2011mr}.
For nucleus-nucleus reactions one observes a rather linear scaling with
$(dN_{ch}/d\eta)^{1/3}$. The scaling is very good if the change in
$(dN_{ch}/d\eta)^{1/3}$ is caused by a change of centrality at a fixed energy.
However, a small offset on the order of 2 fm - 3 fm is visible for different
system sizes, if the radii are extrapolated to $N_{ch}\rightarrow 0$. This is
expected due to the finite size of the nuclei  in AA reactions
\cite{Sinyukov:2011mw}. In contrast, increasing the center-of-mass energy leads 
to a reduction of the radii at a given fixed $N_{ch}$-bin. The scaling of the
source size with $(dN_{ch}/d\eta)^{1/3}$ for different centralities is a hint
that the underlying physics, e.g. pion production via resonance decay versus
production via string fragmentation, is nearly unchanged by changes in the
collision geometry. A change in $\sqrt{s}$ on the other hand results not only
in different weights of the production mechanisms, but also in changed
expansion dynamics towards a more violent expansion with increased energy.
Qualitatively, one expects a scaling of the length of homogeneity as $R =
R_{geom} / \sqrt{1 + \langle v_\perp^2 \rangle m_\perp/2T}$
\cite{Sinyukov:2011mw,Akkelin:1995gh}, where $R_{geom}$ is the geometric size
of the collision region, $v_\perp$ is the transverse flow velocity and T is the
freeze-out temperature. I.e. the increase in transverse flow leads to a decrease of the
observed radii with increasing energy as observed in the model. This
combination leads to a deviation from the $( dN_{ch}/d\eta)^{1/3}$ scaling of
the HBT radii. The proton-proton calculation (and the data) show significantly
smaller radii and a different slope from what is expected from nucleus-nucleus
results. This behaviour is attributed to the strongly different particle
production mechanisms in AA and pp. I.e., bulk emission vs. string/jet
dominated emission which is also in line with the theoretically observed
dependence of the HBT radii on the formation time of the hadrons from the jet
fragmentation and string decay \cite{Graef:2012za}. \\

\begin{figure} 
\center
\includegraphics[width=.8\textwidth]{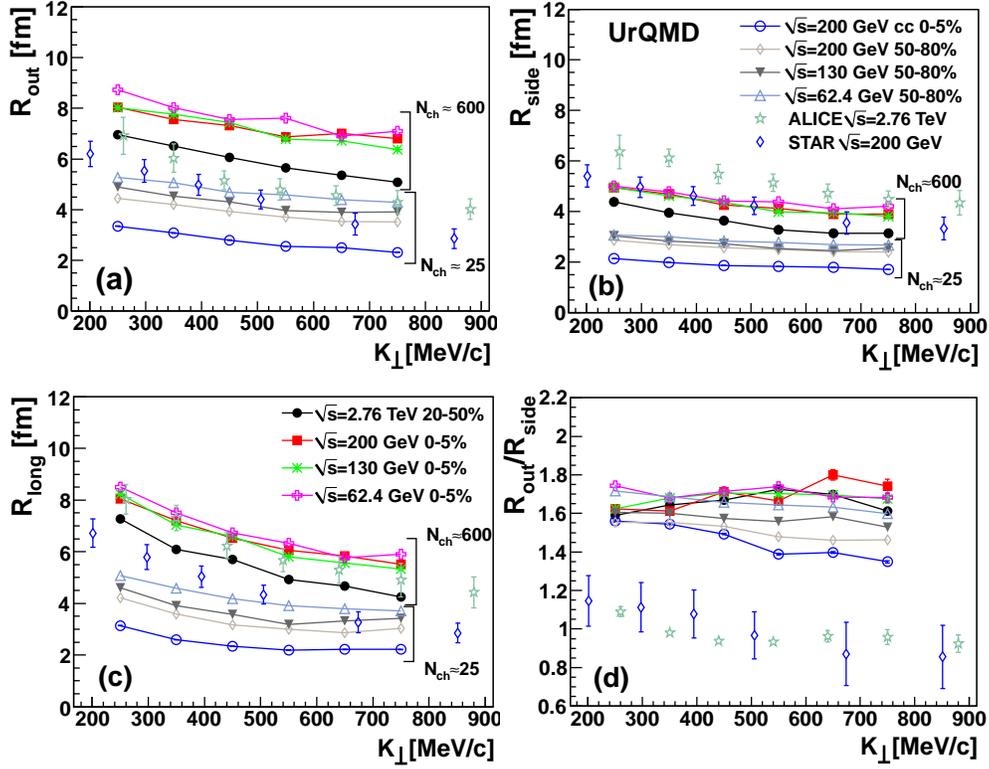}
\caption{(Color online) The $k_\perp$ dependence of $R_{out}$, $R_{side}$ and $R_{long}$. The
black dots are calculations at $\sqrt{s}=$ 2760 GeV and 20-50\% centrality, the
red squares, the green crosses and the pink crosses are lead-lead for 0-5\%
centrality at $\sqrt{s}=$ 200, 130, 62.4 GeV. They have $\left < dN_{ch}/d\eta
\right > \approx$ 670, 665, 595, 509. The other presented calculations are
carbon-carbon at $\sqrt{s}=$ 200 GeV for 0-5\% centrality (blue circles) and
lead-lead at $\sqrt{s}=$ 200, 130, 62.4 GeV (beige diamonds, grey triangles,
blue triangles) all for 50-80\% centrality. These collisions have $\left <
dN_{ch}/d\eta \right > \approx$ 23, 32, 28, 23. The green stars represent ALICE
lead-lead data for central collisions at $\sqrt{s}=$ 2760 GeV
\cite{Aamodt:2011mr}. The blue diamonds are experimental results for central
gold-gold collisions at $\sqrt{s}=$ 200 GeV from the STAR collaborator.
\cite{Adams:2004yc} \label{fig:kt}} 
\end{figure}

Since the $K_\perp$ dependence of the HBT radii tells us much about the
expansion of the source \cite{Pratt:1984su,Hama:1987xv}, let us next
investigate how a variation of $dN_{ch}/d\eta$ is reflected in the differential
HBT radii as recently discussed in \cite{Truesdale:2012zz}. Fig.
\ref{fig:kt} shows the three HBT radii $R_{out}$, $R_{side}$ and $R_{long}$ at
fixed charged particle multiplicity at midrapidity as a function of $k_T$. The
shown calculations are chosen so that they fall roughly into two $\left <
dN_{ch}/d\eta \right >$ classes. The first class contains calculations with
$\left < dN_{ch}/d\eta \right > \approx$ 600 (exact values are 670 for Pb+Pb at
$\sqrt{s}=2760$ GeV, 20-50\% centrality and 665, 595 and 509 for Pb+Pb at
$\sqrt{s}=$ 200, 130, 62.4 GeV, 0-5\% centrality). The second class contains
calculations for $\left < dN_{ch}/d\eta \right > \approx$ 25 (exact values are
23 for C+C at $\sqrt{s} = 200$ GeV, 0-5\% centrality and 32, 28 and 23 for
Pb+Pb at $\sqrt{s}=$ 200, 130, 62.4 GeV and 50-80\% centrality).

A very similar slope in K$_\perp$ is observed for all UrQMD results. This leads
to the conclusion that the observed HBT radii dependence on the radial flow in
the model is weaker than observed in the data. The shift in magnitude of the
radii is related to the magnitude differences already observed in Fig.
\ref{fig:radii} that are mainly dominated by geometry and $\sqrt{s}$ effects.

\section{Volume and freeze-out time}

\begin{figure}
\center
\includegraphics[width=.85\textwidth]{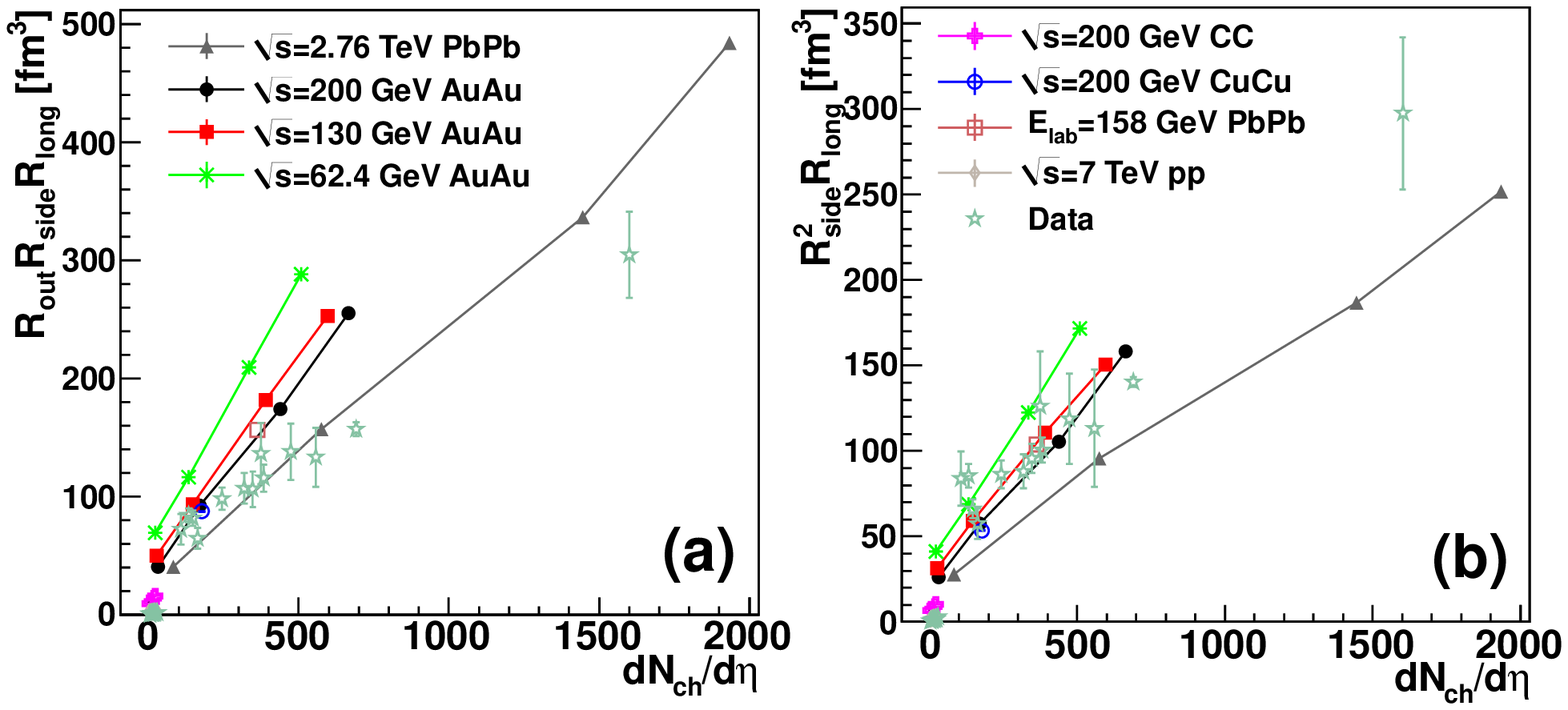}
\caption{(Color online) Two definitions of the volume of homogeneity as a function of energy for various systems. In the left plot the volume is defined as $R_{out} R_{side} R_{long}$ and in the right plot the volume is defined as $R_{side}^2 R_{long}$. The gray triangles, black circles, red squares and green crosses depict UrQMD results for lead-lead collisions at (in this order) $\sqrt{s}=$2760, 200, 130, 62.4 GeV for the centralities 0-5\%, 5-20\%, 20-40\%, 40-80\%. The pink crosses are carbon-carbon calculations at $\sqrt{s}=$ 200 GeV for the same centralities, the blue circles are central copper-copper collisions at $\sqrt{s}=$ 200 GeV and the brown squares are central lead-lead collisions at $E_{lab}=$ 158 AGeV. The beige diamonds depict proton-proton results at $\sqrt{s}=$ 7 TeV for different $(dN_{ch}/d\eta)^{1/3}$ classes. The green stars show experimental results taken from \cite{Adams:2004yc,Lisa:2000no,Alt:2007uj,Afanasiev:2002mx,Adamova:2002wi,Abelev:2009tp,Back:2004ug,Back:2005hs,Back:2002wb,Abelev:2008ez,Aamodt:2011kd,Aamodt:2011mr}. \label{fig:volume}} 
\end{figure}

Next, let us investigate the energy and system size dependence of the
homogeneity volume. Fig. \ref{fig:volume} shows the volume of homogeneity as a
function of $dN_{ch}/d\eta$ for various systems. Lead-lead calculations are
shown for $\sqrt{s}=$ 2760, 200, 130, 62.4 GeV (grey triangles, black circles,
red squares, green crosses) in the centrality classes 0-5\%, 5-20\%, 20-40\%
and 40-80\%. The pink crosses show $\sqrt{s}=$ 200 GeV carbon-carbon results
for the same centralities, and the beige diamonds represent proton-proton
calculations at $\sqrt{s}=$ 7 TeV for  different $dN_{ch}/d\eta$ bins. Blue
circles and brown squares depict results for central copper-copper events at
$\sqrt{s}=$ 200 GeV and central lead-lead events at $E_{lab}=158$ GeV. These
results are compared to experimental data
\cite{Adams:2004yc,Lisa:2000no,Alt:2007uj,Afanasiev:2002mx,Adamova:2002wi,Abelev:2009tp,Back:2004ug,Back:2005hs,Back:2002wb,Abelev:2008ez,Aamodt:2011kd,Aamodt:2011mr}
which is represented by green stars. In line with the experimental data, a
strong increase in the volume proportional to the charged particle multiplicity
is observed. A good agreement between experiment and theory is observed for the
quantity $R_{side}^2R_{long}$ while the experimental results for $R_{out} R_{side} R_{long}$ are
slightly overestimated. This is due to a too large $R_{out}$ in the calculations.
The overestimation of $R_{out}$ is common for hadronic cascade models and can be
explained by a lack of pressure in the early stage of the heavy ion collision
\cite{Li:2007yd,Pratt:2009hu}. While the volume of the homogeneity region for
each individual energy scales very well with $dN_{ch}/d\eta$ Fig.
\ref{fig:volume} shows a steeper slope with decreasing energy. The
calculations also hint to an offset for AA reactions on the order of $25$
fm$^3$ ($R_{side}^2 R_{long}$) and 50 fm$^3$ ($R_{out} R_{side} R_{long}$).\\

\begin{figure}
\center
\includegraphics[width=.7\textwidth]{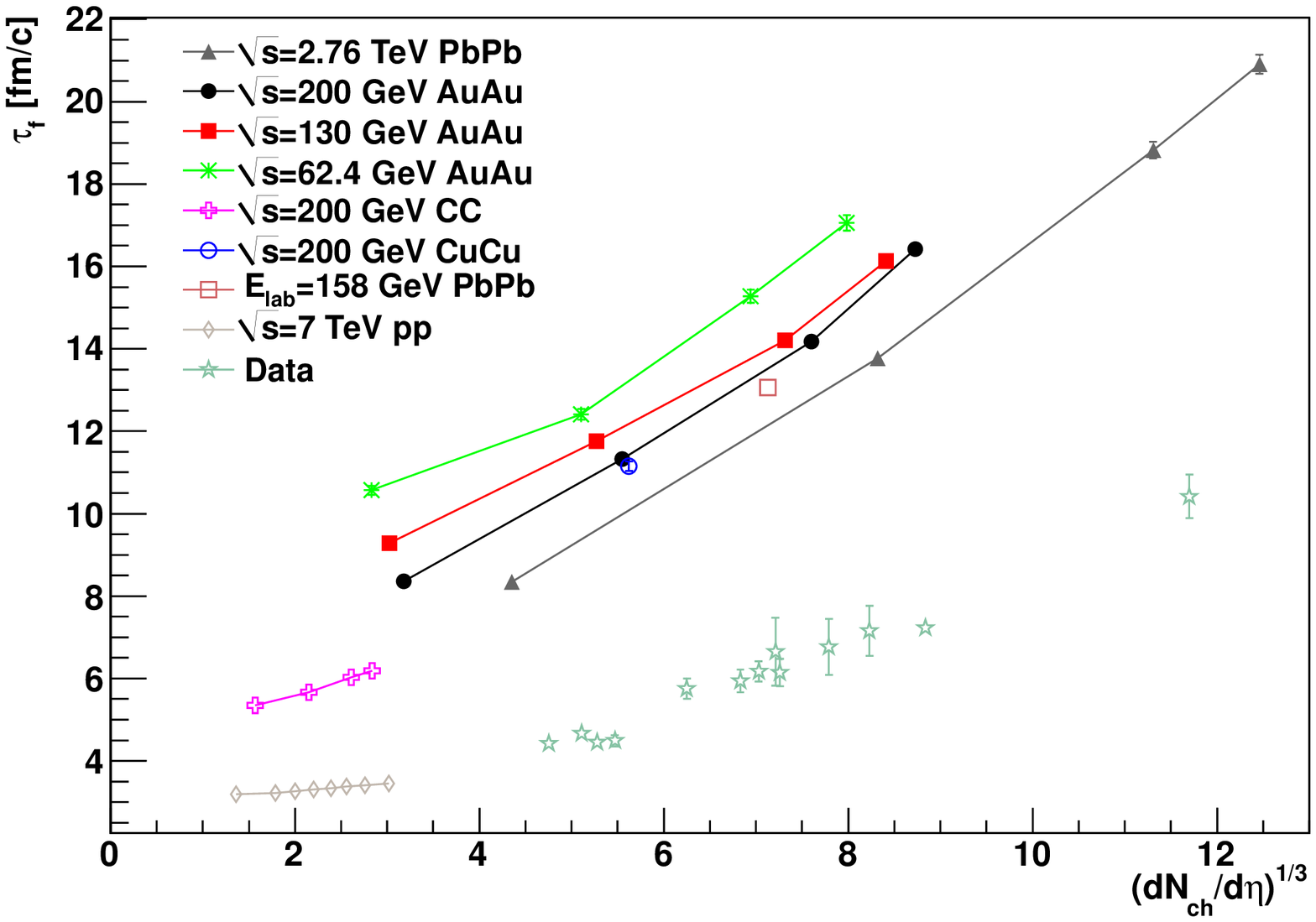}
\caption{(Color online) The freeze-out time as a function of energy for various systems. The
gray triangles, black circles, red squares and green crosses depict UrQMD
results for lead-lead collisions at (in this order) $\sqrt{s}=$2760, 200, 130
62.4 GeV for the centralities 0-5\%, 5-20\%, 20-40\%, 40-80\%. The pink crosses
are carbon-carbon calculations at $\sqrt{s}=$ 200 GeV for the same
centralities, the blue circles are central copper-copper collisions at
$\sqrt{s}=$ 200 GeV and the brown squares are central lead-lead collisions at
$E_{lab}=$ 158 AGeV. The beige diamonds depict proton-proton results at
$\sqrt{s}=$ 7 TeV for different $(dN_{ch}/d\eta)^{1/3}$ classes. The green
stars show experimental results taken from
\cite{Adams:2004yc,Lisa:2000no,Alt:2007uj,Afanasiev:2002mx,Adamova:2002wi,Abelev:2009tp,Back:2004ug,Back:2005hs,Back:2002wb,Abelev:2008ez,Aamodt:2011mr}.
\label{fig:tau}}
\end{figure}

Finally, we explore the apparent freeze-out times $\tau_f$. The results are obtained by fitting the hydrodynamically motivated Eq. \ref{eq:tau} \cite{Aamodt:2011mr,Makhlin:1987gm} to the $k_\perp$ dependence of $R_{long}$ in the interval $K_\perp=$ 200-800 MeV/c. For this purpose the pion freeze-out temperature is assumed to be $T=$ 120 MeV. 
\begin{equation}
  R_{long}^2=\tau_f^2 \frac{T}{m_\perp} \frac{K_2(m_\perp/T)}{K_1(m_\perp/T)},
  \label{eq:tau}
\end{equation}
where $m_\perp = \sqrt{m_\pi^2+k_\perp^2}$ and $K_i$ are the integer order
modified Bessel functions. Fig. \ref{fig:tau} shows the freeze-out time as a
function of $dN_{ch}/d\eta$ for various systems. The grey triangles, the black
circles, the red squares and the green crosses are calculations of lead-lead
collisions at $\sqrt{s}=$ 2760, 200, 130, 62.4 GeV (in the same order) for the
centralities 0-5\%, 5-20\%, 20-40\%, 40-80\%. The pink crosses are
carbon-carbon collisions at $\sqrt{s}=$ 200 GeV for the same centralities. The
blue circles are calculations for central copper-copper collisions at
$\sqrt{s}=$ 200 GeV and central lead-lead collisions at $E_{lab}=$ 158 AGeV.
Experimental results
\cite{Adams:2004yc,Lisa:2000no,Alt:2007uj,Afanasiev:2002mx,Adamova:2002wi,Abelev:2009tp,Back:2004ug,Back:2005hs,Back:2002wb,Abelev:2008ez,Aamodt:2011mr}
are depicted by green stars. 
As for all the other observables, there is scaling for each energy
individually. As anticipated from the calculations of $R_{long}$ the decoupling time
$\tau_f$ increases with decreasing energy. This confirms the idea of a shorter
decoupling time with increased energy. The offset in $\tau_f$ for $dN_{ch}/d\eta
\rightarrow 0 $ seems to hint towards a minimal decoupling time $\tau^{min}_f
\sim 4-8 $fm/c in AA reactions and $\tau_f^{min} < 2$ fm/c in pp.

\section{Summary and outlook}

In the light of recent LHC data on pp and AA collisions, which indicate a
modification of the multiplicity scaling of the HBT radii, we have explored the
$N_{ch}$ scaling for a large variety of systems and energies. We find good
scaling of the radii with $dN_{ch}/d\eta$ within a given system and energy.
While the radii decrease slightly with increasing beam energy, they have a
similar slope when plotted versus $(dN_{ch}/d\eta)^{1/3}$ at all energies. 
When analyzing the freeze-out volume versus $dN_{ch}/d\eta$ the increasing
steepness of the slope for decreasing energies becomes visible. For all
observables the scaling of the results for pp collisions differ strongly from
the nucleus-nucleus results. We relate this observation to the different
particle emission patterns (bulk vs. strings) in AA and pp.

\section*{Acknowledgements} This work was supported by the Helmholtz
International Center for FAIR within the framework of the LOEWE program launched
by the State of Hesse, GSI, and BMBF. G.G. thanks the Helmholtz Research School for Quark Matter Studies (H-QM) for support.
Q.L. thanks the financial support by the key project of the Ministry of Education (No. 209053), the NNSF (Nos. 10905021, 10979023), the Zhejiang Provincial NSF (No. Y6090210), and the Qian-Jiang Talents Project of Zhejiang Province (No. 2010R10102) of China.

\section*{References}


\begin{thebibliography}{36}

\bibitem{Adams:2004yc}
  J.~Adams {\it et al.}  [STAR Collaboration],
  Phys.\ Rev.\  C {\bf 71}, 044906 (2005).

\bibitem{Lisa:2000no}
M.~A. Lisa, et~al., E895 Collaboration, Phys. Rev. Lett. 84 (2000) 2798--2802.

\bibitem{Alt:2007uj}
C.~Alt, et~al., NA49 Collaboration, Phys. Rev. C77 (2008) 064908.

\bibitem{Afanasiev:2002mx}
S.~V. Afanasiev, et~al., NA49 Collaboration, Phys. Rev. C66 (2002) 054902.

\bibitem{Adamova:2002wi}
D.~Adamov\'{a}, et~al., CERES Collaboration, Nucl. Phys. A714 (2003) 124--144.

\bibitem{Abelev:2009tp}
B.~I. Abelev, et~al., STAR Collaboration, Phys. Rev. C80 (2009) 024905.

\bibitem{Back:2004ug}
B.~B. Back, et~al., PHOBOS Collaboration, Phys. Rev. C73 (2006) 031901.

\bibitem{Back:2005hs}
B.~B. Back, et~al., PHOBOS Collaboration, Phys. Rev. C74 (2006) 021901.

\bibitem{Back:2002wb}
B.~B. Back, et~al., PHOBOS Collaboration, Phys. Rev. Lett. 91 (2003) 052303.

\bibitem{Abelev:2008ez}
B.~I. Abelev, et~al., STAR Collaboration, Phys. Rev. C79 (2009) 034909.

\bibitem{Aamodt:2011kd}
  K.~Aamodt {\it et al.} [ ALICE Collaboration ],
  [arXiv:1101.3665 [hep-ex]].

\bibitem{Aamodt:2011mr}
  K.~Aamodt {\it et al.}  [ALICE Collaboration],
  Phys.\ Lett.\  B {\bf 696}, 328 (2011).

\bibitem{Pratt:1984su} 
  S.~Pratt,
  Phys.\ Rev.\ Lett.\  {\bf 53}, 1219 (1984).

\bibitem{Sinyukov:1989xz} 
  Y.~.M.~Sinyukov,
  Nucl.\ Phys.\ A {\bf 498}, 151C (1989).

\bibitem{Hama:1987xv} 
  Y.~Hama and S.~S.~Padula,
  Phys.\ Rev.\ D {\bf 37}, 3237 (1988).

\bibitem{nucl-th/9901094} 
  U.~A.~Wiedemann and U.~W.~Heinz,
  Phys.\ Rept.\ \ {\bf 319}, 145  (1999).

\bibitem{Lisa:2005dd} 
  M.~A.~Lisa, S.~Pratt, R.~Soltz and U.~Wiedemann,
  Ann.\ Rev.\ Nucl.\ Part.\ Sci.\  {\bf 55}, 357 (2005).

\bibitem{Sinyukov:2011mw} 
  Y.~.M.~Sinyukov and I.~.A.~Karpenko,
  Phys.\ Part.\ Nucl.\ Lett.\  {\bf 8}, 896 (2011).

\bibitem{Akkelin:1995gh} 
  S.~V.~Akkelin and Y.~.M.~Sinyukov,
  Phys.\ Lett.\ B {\bf 356}, 525 (1995).


\bibitem{Adamova:2002ff} 
  D.~Adamova {\it et al.}  [CERES Collaboration],
  Phys.\ Rev.\ Lett.\  {\bf 90}, 022301 (2003).

\bibitem{Akkelin:2003kp} 
  S.~V.~Akkelin and Y.~.M.~Sinyukov,
  nucl-th/0310036.

\bibitem{Akkelin:2005ms} 
  S.~V.~Akkelin and Y.~.M.~Sinyukov,
  Phys.\ Rev.\ C {\bf 73}, 034908 (2006).

\bibitem{Graef:2012za} 
  G.~Graef, Q.~Li and M.~Bleicher,
  arXiv:1203.4421 [nucl-th].

\bibitem{Li:2012ta} 
  Q.~Li, G.~Graef and M.~Bleicher,
  arXiv:1203.4104 [nucl-th].

\bibitem{Bass:1998ca}
  S.~A.~Bass, M.~Belkacem, M.~Bleicher, M.~Brandstetter, L.~Bravina, C.~Ernst, L.~Gerland, M.~Hofmann {\it et al.},
  Prog.\ Part.\ Nucl.\ Phys.\  {\bf 41}, 255-369 (1998).

\bibitem{Bleicher:1999xi}
  M.~Bleicher, E.~Zabrodin, C.~Spieles, S.~A.~Bass, C.~Ernst, S.~Soff, L.~Bravina, M.~Belkacem {\it et al.},
  J.\ Phys.\ G {\bf G25}, 1859-1896 (1999).

\bibitem{Petersen:2008kb}
  H.~Petersen, M.~Bleicher, S.~A.~Bass, H.~Stocker,
  [arXiv:0805.0567 [hep-ph]].

\bibitem{Petersen:2008dd}
  H.~Petersen, J.~Steinheimer, G.~Burau, M.~Bleicher, H.~Stocker,
  Phys.\ Rev.\  {\bf C78}, 044901 (2008).

\bibitem{urqmd-webpage}
Download the most recent UrQMD source code from http://urqmd.org/

\bibitem{Li:2006gp} 
  Q.~Li, M.~Bleicher and H.~Stoecker,
  Phys.\ Rev.\ C {\bf 73}, 064908 (2006).

\bibitem{Li:2007yd} 
  Q.~Li, M.~Bleicher and H.~Stocker,
  Phys.\ Lett.\ B {\bf 659}, 525 (2008).

\bibitem{Li:2008qm} 
  Q.~Li, J.~Steinheimer, H.~Petersen, M.~Bleicher and H.~Stocker,
  Phys.\ Lett.\ B {\bf 674}, 111 (2009).

\bibitem{Li:2008ge} 
  Q.~Li, M.~Bleicher and H.~Stocker,
  Phys.\ Lett.\ B {\bf 663}, 395 (2008).

\bibitem{Truesdale:2012zz} 
  D.~Truesdale and T.~J.~Humanic,
  J.\ Phys.\ G G {\bf 39}, 015011 (2012).

\bibitem{Pratt:2009hu} 
  S.~Pratt,
  Nucl.\ Phys.\ A {\bf 830}, 51C (2009).

\bibitem{Makhlin:1987gm}
  A.~N.~Makhlin, Y.~.M.~Sinyukov,
  Z.\ Phys.\  {\bf C39}, 69 (1988).
  
\end{thebibliography}
\end{document}